\documentclass[twocolumn]{aastex62}
\NewPageAfterKeywords
\usepackage[hyphenbreaks]{breakurl}
\usepackage{longtable}
\usepackage{threeparttablex}
\usepackage{comment}
\usepackage{lineno}

\newcommand{\newcompanions}{26}

\newcommand{\knownmultiples}{290}
\newcommand{\pokemonnum}{1070}
\newcommand{\percentincrease}{7.6\%} 
\newcommand{\gaiaparallax}{17}

\newcommand{\otherparallax}{nine}

\graphicspath{{./}{figures/}}

\received{2022 April 7}
\revised{2022 May 5}
\accepted{2022 May 24}
\submitjournal{AJ}

\begin{document}

\title{The POKEMON Speckle Survey of Nearby M Dwarfs. I. New Discoveries}

\correspondingauthor{Catherine A. Clark}
\email{catclark@nau.edu}

\author[0000-0002-2361-5812]{Catherine A. Clark}
\affil{Northern Arizona University, 527 South Beaver Street, Flagstaff, AZ 86011, USA}
\affil{Lowell Observatory, 1400 West Mars Hill Road, Flagstaff, AZ 86001, USA}

\author[0000-0002-8552-158X]{Gerard T. van Belle}
\affil{Lowell Observatory, 1400 West Mars Hill Road, Flagstaff, AZ 86001, USA}

\author[0000-0003-2159-1463]{Elliott P. Horch}
D\affil{Southern Connecticut State University, 501 Crescent Street, New Haven, CT 06515, USA}

\author[0000-0002-5823-4630]{Kaspar von Braun}
\affil{Lowell Observatory, 1400 West Mars Hill Road, Flagstaff, AZ 86001, USA}

\author[0000-0002-5741-3047]{David R. Ciardi}
\affil{NASA Exoplanet Science Institute Caltech/IPAC, Pasadena, CA 91125, USA}


\author[0000-0001-6031-9513]{Jennifer G. Winters}
\affil{Center for Astrophysics $\vert$ Harvard \& Smithsonian, 60 Garden Street, Cambridge, MA 02138, USA}


\author[0000-0003-2102-3159]{Rocio Kiman}
\affil{Kavli Institute for Theoretical Physics, University of California, Santa Barbara, CA 93106, USA}



\begin{abstract}

M dwarfs are favorable targets for exoplanet detection with current instrumentation, but stellar companions can induce false positives and inhibit planet characterization. Knowledge of stellar companions is also critical to our understanding of how binary stars form and evolve. We have therefore conducted a survey of stellar companions around nearby M dwarfs, and here we present our new discoveries. Using the DSSI speckle imager at the 4.3-meter Lowell Discovery Telescope, and the similar NESSI instrument at the 3.5-meter WIYN telescope, we carried out a volume-limited survey of M-dwarf multiplicity to 15 parsecs, with a special emphasis on including the later M dwarfs that were overlooked in previous surveys. Additional brighter targets at larger distances were included for a total sample size of \pokemonnum{} M dwarfs. Observations of these \pokemonnum{} targets revealed \newcompanions{} new companions; 22 of these systems were previously thought to be single. If all new discoveries are confirmed, then the number of known multiples in the sample will increase by \percentincrease{}. Using our observed properties, as well as the parallaxes and 2MASS $K$ magnitudes for these objects, we calculate the projected separation, and estimate the mass ratio and component spectral types, for these systems. We report the discovery of a new M-dwarf companion to the white dwarf Wolf 672 A, which hosts a known M-dwarf companion as well, making the system trinary. We also examine the possibility that the new companion to 2MASS J13092185-2330350 is a brown dwarf. Finally, we discuss initial insights from the POKEMON survey.

\end{abstract}



\section{Introduction} \label{sec:intro}


Though they are the smallest and least luminous stars on the Main Sequence, the M dwarfs occupy a captivating range of stellar parameter space. Their masses span nearly a factor of ten \citep{Baraffe1996ApJ...461L..51B}, and the lowest-mass M dwarfs have Main Sequence lifetimes of trillions of years \citep{Laughlin1997ApJ...482..420L}. Additionally, the M dwarfs dominate the galactic neighborhood, accounting for approximately 75\% of the stars in the Milky Way \citep{Henry2006AJ....132.2360H}. The M dwarfs also offer a significant opportunity for finding and characterizing Earth-sized planets \citep[e.g.,][]{Lopez-Morales2019AJ....158...24L}, due to their relatively small star-to-planet mass and radius ratios, as well as their sheer numbers.

However, one property that can inhibit the detection and characterization of the planets that orbit M dwarfs is stellar multiplicity. 
``Third light'' contamination of light curves, which is caused by stellar companions, has been shown to inhibit the detection of Earth-sized, transiting planets \citep{Lester2021AJ....162...75L}, and has led to additional obstacles in planet characterization including underestimated planet radii \citep{Ciardi2015ApJ...805...16C}, skewed planet radius distributions and occurrence rates \citep{Hirsch2017AJ....153..117H, Teske2018AJ....156..292T, Bouma2018AJ....155..244B}, incorrect characterization of both stars' properties \citep{FurlanHowell2020ApJ...898...47F}, and improper mean density and atmospheric values \citep{Howell2020FrASS...7...10H}. Additionally, close-in stellar companions can perturb and truncate protoplanetary disks \citep{Jang-Condell2015ApJ...799..147J}, gravitationally excite planetesimals causing collisional destruction \citep{RafikovSilsbee2015aApJ...798...69R, RafikovSilsbee2015bApJ...798...70R}, and scatter and eject planets that have formed \citep{HaghighipourRaymond2007ApJ...666..436H}. Furthermore, recent studies have suggested that even wide stellar companions might affect the formation or orbital properties of giant planets \citep[e.g.][]{Fontanive2019MNRAS.485.4967F, FontaniveBardalezGagliuffi2021FrASS...8...16F, Hirsch2021AJ....161..134H, Mustill2022A&A...658A.199M, Su2021AJ....162..272S}. As the M dwarfs have proven to be such favorable targets for planet detection and characterization, measuring their multiplicity is therefore crucial to understanding the planets that they host.

Furthermore, work on Kepler \citep{Borucki2011ApJ...728..117B}, K2 \citep{Howell2014PASP..126..398H}, and now TESS \citep{Ricker2015JATIS...1a4003R} suggests that the stellar companions to exoplanet hosts have longer orbital periods than the companions to field stars \citep[e.g.,][]{Kraus2012ApJ...745...19K, Bergfors2013MNRAS.428..182B, Wang2014ApJ...791..111W, Kraus2016AJ....152....8K, Ziegler2020AJ....159...19Z, Hirsch2021AJ....161..134H, Howell2021AJ....161..164H, Lester2021AJ....162...75L, MoeKratter2021MNRAS.507.3593M, Clark2022AJ....163..232C}. Multiplicity measurements are therefore critical to understanding M-dwarf system architectures and occurrence rates as well.

Because of the significance M dwarfs have for both stellar astrophysics and exoplanet studies, many surveys have been carried out to determine the M-dwarf multiplicity rate, especially in recent years. It has been shown that multiplicity decreases with mass from O to M \citep[e.g.,][]{Mason2009AJ....137.3358M, Raghavan2010ApJS..190....1R, Duchene2013ARA&A..51..269D, Winters2019AJ....157..216W}, but this multiplicity has not been fully characterized for M dwarfs, particularly for the later sub-types, due to their faintness and to resolution limits. This is highlighted by the fact that a reasonably complete inventory of later M dwarfs did not exist until recently \citep[e.g.,][]{Kirkpatrick2014ApJ...783..122K,LuhmanSheppard2014ApJ...787..126L,Winters2021AJ....161...63W}. \citet{Winters2019AJ....157..216W} produced the most comprehensive M-dwarf multiplicity study to date: an all-sky, volume-limited survey that extends to 25 pc. However, 
their study surveyed binaries with separations larger than $2\arcsec$. The study presented here therefore complements the \citet{Winters2019AJ....157..216W} survey by exploring the inner regions around nearby M dwarfs with high-resolution imaging. 

Here we present the \newcompanions{} new discoveries detected throughout the Pervasive Overview of `Kompanions' of Every M dwarf in Our Neighborhood (POKEMON) survey, which identified companions to nearby M dwarfs at large-telescope, diffraction-limited resolution. The POKEMON survey is volume-limited through M9, out to at least 15 pc, with additional brighter targets at larger distances, resulting in a sample of \pokemonnum{} nearby M dwarfs. A future paper will present the full sample of M dwarfs we surveyed, the completeness of the survey, and our updated M-dwarf multiplicity rate. Additionally, though \citet{Janson2012ApJ...754...44J} calculated M-dwarf multiplicity by spectral sub-type through M6, our upcoming paper will establish the M-dwarf multiplicity rate by sub-type through M9 for the first time. This initial paper in the series presents and characterizes the new discoveries that have been revealed by the POKEMON survey.

In Section \ref{sec:observations}, we describe our target selection process, observational routine, and data reduction procedure. In Section \ref{sec:results}, we note systems with both a new discovery and a known companion, and we provide observed properties for the systems with new discoveries. We also assess the likelihood that these new discoveries are bound. Additionally, we estimate astrophysical properties for the systems with new discoveries using the observed properties and parallaxes from the literature. In Section \ref{sec:discussion}, we discuss a system with both a known M dwarf and a known white dwarf, a system with a potential brown dwarf companion, and initial insights from the POKEMON survey. We summarize our conclusions and discuss future work in Section \ref{sec:conclusions}.

\clearpage

\section{Observations} \label{sec:observations}

The POKEMON survey used speckle interferometry \citep{Labeyrie1970A&A.....6...85L}, also known as speckle imaging, to observe \pokemonnum{} M dwarfs throughout the northern sky.

\subsection{Target Selection}

The initial basis for our target selection was the seminal Catalog of Nearby Stars (CNS3; \citealt{Gliese1991adc..rept.....G}), with updates from recent nearby neighbor discoveries from surveys such as RECONS \citep{Henry2006AJ....132.2360H, Winters2015AJ....149....5W} and 2MASS \citep{Skrutskie2006AJ....131.1163S}. A preliminary review of these augmented CNS3 data indicated 522 objects out to 15 pc either identified specifically as M dwarfs, or as potential M dwarfs from $H - K$ color and absolute magnitude $M_V$. Out to 25 pc, this review resulted in approximately 1,500 objects.

We expanded this list to include the AllWISE Motion Survey \citep{Kirkpatrick2014ApJ...783..122K} with 3,525 new high proper motion objects, and the similar (but non-overlapping) study by \citet{LuhmanSheppard2014ApJ...787..126L} that identified 762 high proper motion objects. An examination of the former’s new high proper motion objects, using the spectral type-color relationships from \citet{LuhmanSheppard2014ApJ...787..126L}, indicated that there were $\sim350$ additional objects between the spectral types of M4 and M9. Other surveys that were mined included the Database of Ultracool Parallaxes based on the Hawaii Infrared Parallax Program \citep{DupuyLiu2012ApJS..201...19D}, the CARMENES and APOGEE input catalogs \citep{Alonso-Floriano2015A&A...577A.128A, Deshpande2013AJ....146..156D}, and the 2016 release of the Pan-STARRS Parallax and Proper Motion Catalog \citep{Waters2015IAUGA..2256019W}, which already had a specific focus on identifying nearby low-mass stars \citep{Magnier2015IAUGA..2257922M}.

Given our use of northern hemisphere facilities, a declination cut of $\delta > -30^o$ was carried out. A cut at 15.5 in the $I$ band was also necessary for most objects due to the faint limit of the instruments. These cuts resulted in a sample consisting of \pokemonnum{} objects.

We note that the parallax sources we used to create the original sample were not perfect; newer astrometric data from Gaia \citep{Gaia2018yCat.1345....0G, Gaia2020yCat.1350....0G} have now provided us with more accurate and precise parallax measurements that indicate some stars in our sample are at distances larger than 15 pc. We chose to keep these stars in the sample despite being farther than previously thought. Parallaxes for these objects are discussed further in Section \ref{subsec:likelihood}.

\begin{figure*}
    \centering
    \includegraphics[width=0.9\textwidth]{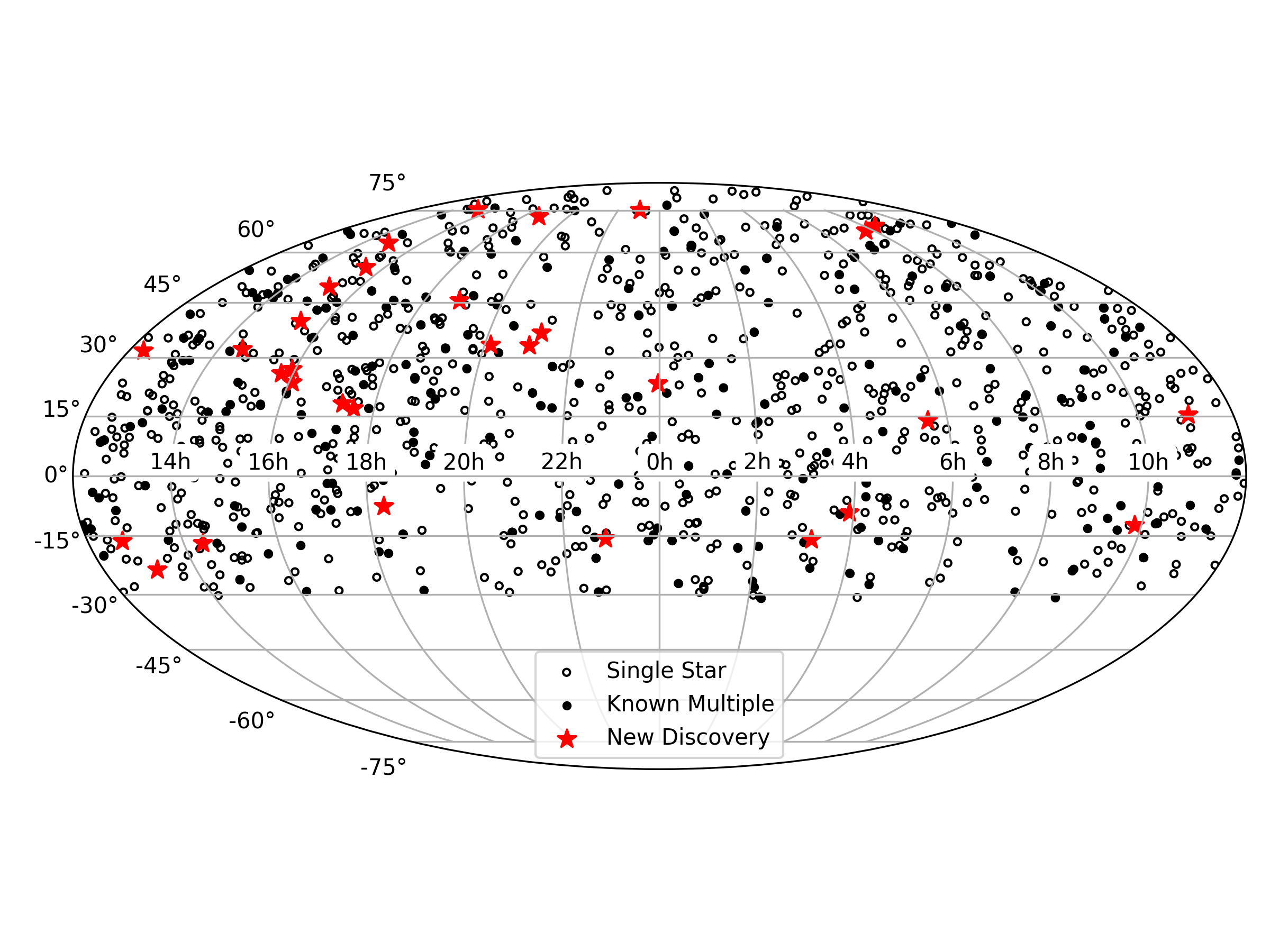}
    \caption{The sky locations of the \pokemonnum{} stars in the POKEMON sample. \knownmultiples{} of these are known multiples. Single stars are marked with open black circles, known multiples are marked with filled black circles, and the \newcompanions{} systems with new discoveries are marked with larger, red stars.}
    \label{fig:aitoff}
\end{figure*}

We present our sample in Figure \ref{fig:aitoff}, which is an Aitoff projection showing sky locations of the single and multiple stars in the full sample, compared with the sky locations of the \newcompanions{} new companions.

\subsection{Observational Routine} \label{subsec:observational_routine}

We imaged the \pokemonnum{} M dwarfs in the POKEMON sample over 50 nights between UT 2017 April 7 and UT 2020 February 10. The main instrument used in the survey was the Differential Speckle Survey Instrument \citep[DSSI;][]{Horch2009AJ....137.5057H}, which was resident at the 4.3-meter Lowell Discovery Telescope (LDT) in Happy Jack, AZ, throughout the POKEMON survey. For brighter targets ($V < 11$), we used the NN-EXPLORE Exoplanet Stellar Speckle Imager \citep[NESSI;][]{Scott2018PASP..130e4502S}, which was used on the 3.5-meter WIYN telescope at Kitt Peak National Observatory outside Tucson, AZ.

DSSI and NESSI produce diffraction-limited images from speckle patterns observed simultaneously at two wavelengths. Each instrument uses a dichroic filter to split the collimated beam from the telescope at $\sim$700 nm into two channels that are imaged on separate high-speed readout EMCCDs. The standard filter arrangement is 692 and 880 nanometers (nm) for DSSI, and 562 and 832 nm for NESSI, each with 40, 44, or 50 nm bandpasses. The limiting spatial resolution of these instruments at the telescopes we used is $\sim 40$ milliarcseconds (mas), which is comparable to the near-infrared adaptive optics observing on the Keck II Telescope. These speckle cameras can therefore identify stellar companions down to the $\sim 1$ au scale.

Bright objects ($V<11$) require only $\sim 1-2$ minutes of observing time, during which the speckle camera obtains a single image cube of 1,000 40-millisecond speckle frames; these short exposures are necessary to ``freeze'' out the atmosphere in our observations and to obtain good speckle contrast. Fainter stars ($11<V<15.5$) require up to $\sim 10$ minutes of observing time per target, during which the speckle camera obtains up to nine image cubes. Standard observing also includes periodic observations of bright, unresolved, single stars from the Bright Star Catalog \citep{HoffleitJaschek1982bsc..book.....H} to probe the atmospheric conditions experienced by the target of interest. All data cubes are stored as multi-extension FITS files.

The pixel scale and image orientation are empirically confirmed by observing binaries with extremely well-known orbits \citep[those listed as Grade 1 in the Sixth Orbit Catalog;][]{Hartkopf2001AJ....122.3472H}. Their ephemeris positions are computed based on the orbital elements, and their scale and orientation are derived from these results. We assume that the calibrations from \citet{Horch2021AJ....161..295H} are also appropriate for this work, given that our targets were observed on the same runs, besides those observed in February 2020. Figure 3 of \citet{Horch2021AJ....161..295H} shows the derived uncertainties from the known orbital elements that we have incorporated into our analysis. However, \citet{Horch2021AJ....161..295H} did not include data from February 2020. To reduce and analyze this data, we used pixel scale values that were approximate, and that may be updated in the future; the position angle and angular separation values from this run may therefore be updated in the future as well.

It should be noted during our first observing run, from UT 2017 April 7 to UT 2017 April 17, the narrowband speckle filters were not installed within DSSI. When observing without filters, more light reaches the detector, but there is less contrast in the speckles. Furthermore, when filters are not used, significant chromatic effects arise at higher airmasses. We did find that the majority of our newly detected companions were first, or only, detected during this observing run. However, we returned to using the speckle filters for all other observing runs in order to increase the contrast in the speckles. Nonetheless, we plan to investigate the possibility of observing without speckle filters further in the future, both to confirm companions observed only during a single epoch, and to probe new speckle imaging discovery space. The most unique discovery from our filterless observations is discussed further in Section \ref{subsec:brown_dwarf}.

Our observations of the new companions are summarized in Table \ref{table:observations}. The 2MASS ID, common name or identifier (if applicable), UT date, telescope, bandpass ($\lambda$), and bandpass width ($\Delta\lambda$) are listed. For the filterless observations, instead of listing the bandpass and bandpass width of the observation, we instead indicate whether the companion was detected in the ``blue'' ($\lambda\lessapprox700$ nm) image or the ``red'' ($\lambda\gtrapprox700$ nm) image. We note that many of the new discoveries were only detected in the ``red'' image, since the M dwarfs emit most strongly at near-infrared wavelengths.

\begin{deluxetable*}{ccccrc}
\tablecaption{Summary of observations for targets with a new companion
\label{table:observations}}
\tablehead{\colhead{2MASS ID} & \colhead{Name} & \colhead{UT Date} & \colhead{Telescope} & \colhead{$\lambda$} & \colhead{$\Delta\lambda$} \\ 
\colhead{} & \colhead{} & \colhead{(YYYY-MM-DD)} & \colhead{} & \colhead{(nm)} & \colhead{(nm)}}
\startdata
 $03104962-1549408$ & LP 772-11 & 2017-10-20 & LDT & 880 & 50 \\ 
 $03542561-0909316$ & & 2017-10-21 & LDT & 880 & 50 \\
 $07011725+1348085$ & G 110-29 & 2017-04-15 & LDT & red & \\
 $07411976+6718444$ & LP 58-260 & 2017-04-17 & LDT & red & \\
 $09510964-1219478$ & GJ 369 & 2017-05-07 & LDT & 880 & 50 \\
 $11030845+1517518$ & L 1258-55 & 2017-04-17 & LDT & blue & \\
 & & 2017-04-17 & LDT & red & \\
 & & 2018-01-31 & LDT & 692 & 40 \\
 & & 2018-01-31 & LDT & 880 & 50 \\
 $12190600+3150433$ & LTT 13435 & 2017-04-17 & LDT & blue & \\
 & & 2017-04-17 & LDT & red & \\
 & & 2018-01-31 & LDT & 692 & 40 \\
 & & 2018-01-31 & LDT & 880 & 50 \\
 $12435889-1614351$ & LP 796-1 & 2017-04-15 & LDT & red & \\
 $13092185-2330350$ & CE 303 & 2017-04-16 & LDT & blue & \\
 & & 2017-04-16 & LDT & red & \\
 $14235017-1646116$ & & 2017-04-17 & LDT & red & \\
 & & 2018-01-31 & LDT & 880 & 50 \\
 $15020759+7527526$ & LP 22-174 & 2017-04-13 & LDT & red & \\
 & & 2019-09-14 & LDT & 880 & 50 \\
 $15085332+4934062$ & & 2017-04-13 & LDT & red & \\
 & & 2020-02-09 & LDT & 880 & 50 \\
 $15211607+3945164$ & LP 222-70 & 2017-04-07 & LDT & red & \\
 $15263317+5522206$ & LP 135-316 & 2017-04-07 & LDT & red & \\
 $15434848+2552376$ & G 167-54 & 2017-04-12 & LDT & blue & \\
 & & 2017-04-12 & LDT & red & \\
 $15471513+0149218$ & LP 623-40 & 2017-04-15 & LDT & red & \\
 $16041322+2331386$ & & 2017-04-12 & LDT & blue & \\
 & & 2017-04-12 & LDT & red & \\
 & & 2017-05-04 & LDT & red & \\
 $17183572+0156433$ & Wolf 672 B & 2017-04-09 & LDT & blue & \\
 & & 2017-04-09 & LDT & red & \\
 $17335314+1655129$ & & 2017-04-17 & LDT & red & \\
 & & 2019-09-13 & LDT & 880 & 50 \\
 $18191622-0734518$ & & 2017-04-11 & LDT & red & \\
 $18523373+4538317$ & LHS 3420 & 2017-04-12 & LDT & red & \\
 & & 2019-09-13 & LDT & 880 & 50 \\
 $20081786+3318122$ & GJ 1250 & 2019-09-14 & LDT & 880 & 50 \\
 $21011610+3314328$ & L 1504-143 & 2017-10-21 & LDT & 692 & 40 \\
 & & 2017-10-21 & LDT & 880 & 50 \\
 & & 2018-08-02 & WIYN & 832 & 40 \\
 $21134479+3634517$ & & 2018-08-29 & WIYN & 832 & 40 \\
 & & 2019-09-13 & LDT & 880 & 50 \\
 $22520522-1532511$ & LP 821-27 & 2019-09-14 & LDT & 880 & 50 \\
 $23024353+7505591$ & LP 49-357 & 2019-01-20 & WIYN & 562 & 44 \\
 & & 2019-01-20 & WIYN & 832 & 40 \\
 \enddata
\end{deluxetable*}

\begin{figure*}[t]
     \centering
     \includegraphics[width=0.49\textwidth]{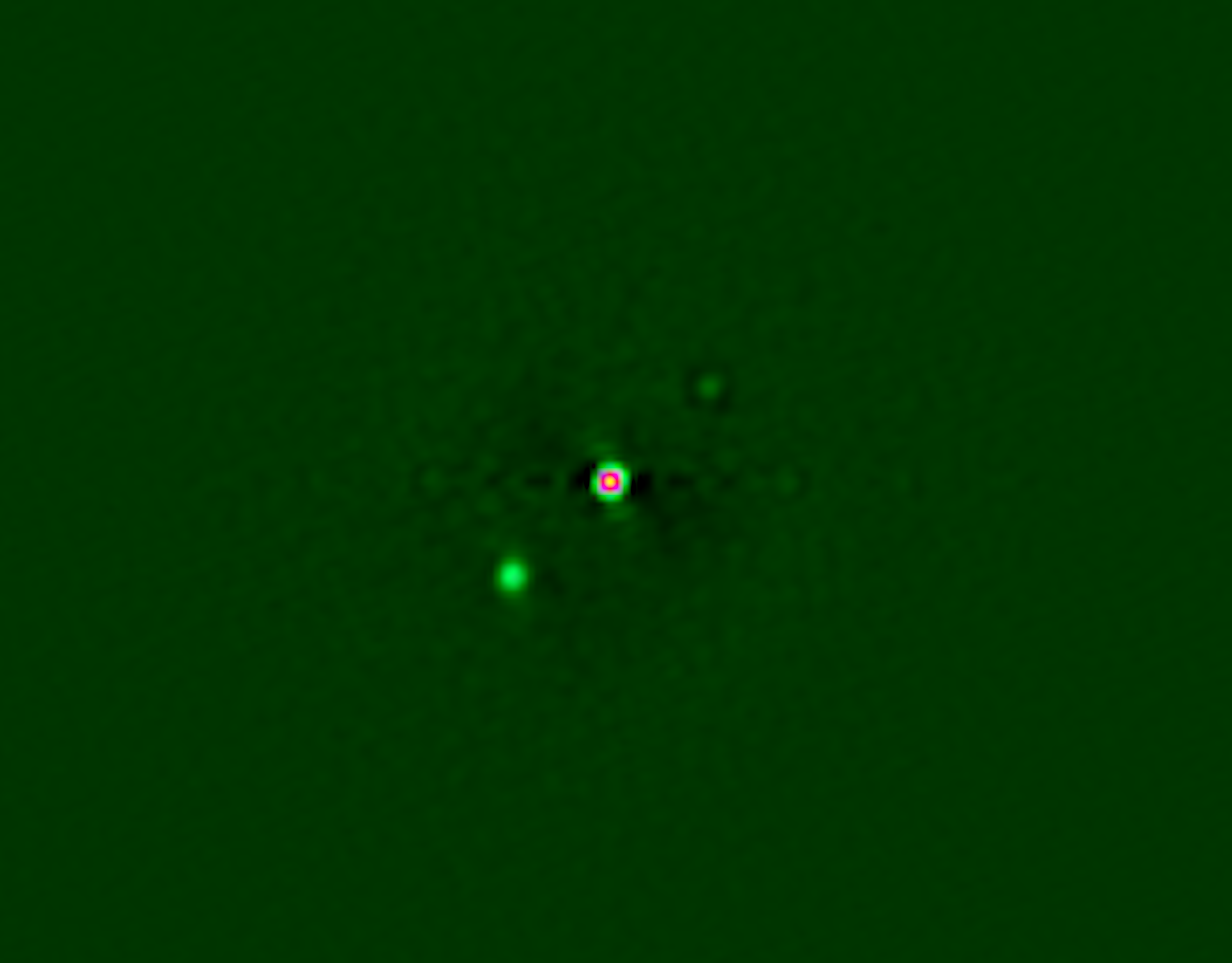}
     \includegraphics[width=0.49\textwidth]{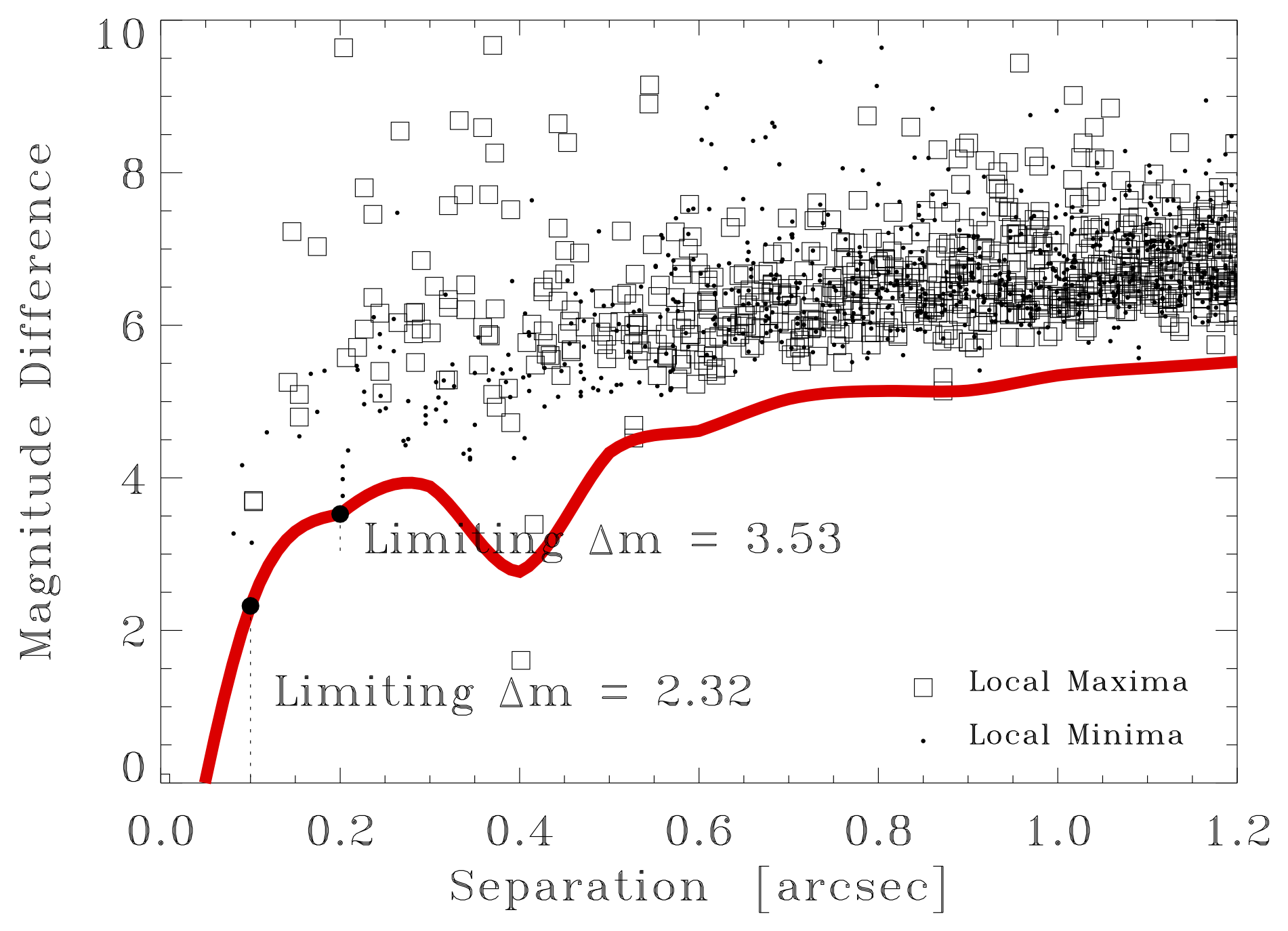}
     \caption{The newly detected companion to 2MASS J11030845+1517518, shown in the reconstructed image on the left and in the detection plot on the right. These data were taken on UT 2017 April 17. The companion is at a separation of $0.39\arcsec$ (6.8 au), and its delta magnitude is measured to be 1.5.}
     \label{fig:example_data}
\end{figure*}

\subsection{Data Reduction}

The data are reduced with a version of the bispectrum speckle reduction code described in \citet{Horch2009AJ....137.5057H,Horch2011aAJ....141...45H, Horch2011bAJ....141..180H}, which uses bispectral analysis \citep{Lohmann1983ApOpt..22.4028L} to compute a reconstructed image. This code uses the fact that binary systems produce a fringe pattern in the Fourier plane. A 2-D autocorrelation function is calculated for each speckle frame and summed over all frames. The Fourier transform of the autocorrelation function is found and squared to obtain the power spectrum, which is normalized. After dividing by the power spectrum of a point source, the residual 2-D power spectrum appears as a set of fringes for each pair of stars in the field. This is fit using a cosine-squared function to determine the relative astrometry and photometry (position angle, angular separation, and delta magnitude) of any pairs of stars in the field. Reconstructed images are constructed from the object's modulus (the square-root of the power spectrum), and the phase estimate is obtained from the bispectrum.

By examining annuli in the reconstructed image that are centered on the primary star, we determine all local maxima and minima in the annulus, and derive their mean value and standard deviation. We then estimate the detection limit as the mean value of the maxima plus five times the average sigma of the maxima and minima. In doing so, the data reduction pipeline produces a curve of this detection limit as a function of separation. Example data demonstrating the reconstructed image and contrast curve are shown in Figure \ref{fig:example_data}.

\section{Results} \label{sec:results}

Our observations of \pokemonnum{} nearby M dwarfs revealed 145 known multiples and \newcompanions{} previously undetected stellar companions. A future paper will present our detections of known multiples; here we present the new discoveries. In this section, we discuss systems that host both a new discovery and a known companion. We also report observed and astrophysical properties for the new discoveries; the observed properties were measured via our speckle observations, and the astrophysical properties were estimated using the observed properties as well as parallaxes from the literature.

\subsection{Known Companions}

Four of the \newcompanions{} systems with new discoveries host known companions from the literature as well. This means that these systems are now trinary, or in the case of 2MASS J21011610+3314328, quaternary, as the B component of the system (2MASS J21012062+3314280) hosts a close-in stellar companion as well \citep{Janson2014ApJS..214...17J}. Properties for the four systems with known companions are included in Table \ref{table:known_companions}.

In the case of 2MASS J17183572+0156433, the known companion is a white dwarf. This system is discussed in detail in Section \ref{subsec:white_dwarf}.

\begin{deluxetable*}{c c c c c}[h]
\tablecaption{Properties for systems with a previously known companion
\label{table:known_companions}}
\tablehead{\colhead{2MASS ID} & \colhead{$\theta$} & \colhead{$\rho$} & \colhead{$\Delta$m} & \colhead{Reference} \\ 
\colhead{} & \colhead{($^{\circ}$)} & \colhead{($\arcsec$)} & \colhead{} & \colhead{}}
\startdata
 $12190600+3150433$ & 222 & 1.7 & 2.9 & \citet{Lamman2020AJ....159..139L} \\
 $17183572+0156433$ & 139 & 13 & 0.1 & \citet{Luyten(1997)} \\
 $21011610+3314328$ & 95 & 57 & 0.9 & \citet{Luyten(1997)} \\
 $23024353+7505591$ & 209 & 4.0 & 1.3 & \citet{Luyten(1997)} \\
 \enddata
\end{deluxetable*}

\subsection{New Discoveries} \label{subsec:new_discoveries}

\begin{figure*}
    \centering
    \includegraphics[width=0.49\textwidth]{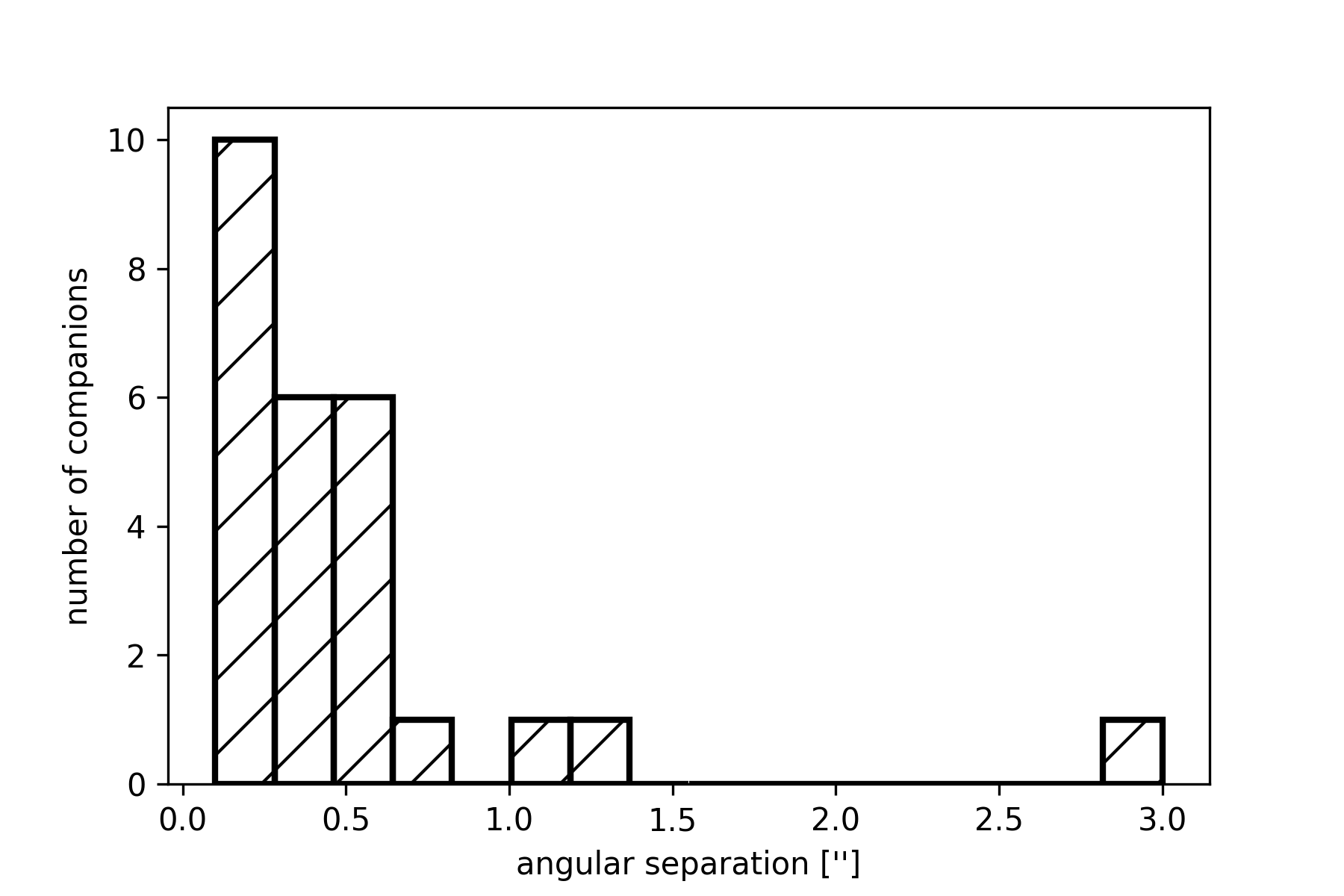}
    \includegraphics[width=0.49\textwidth]{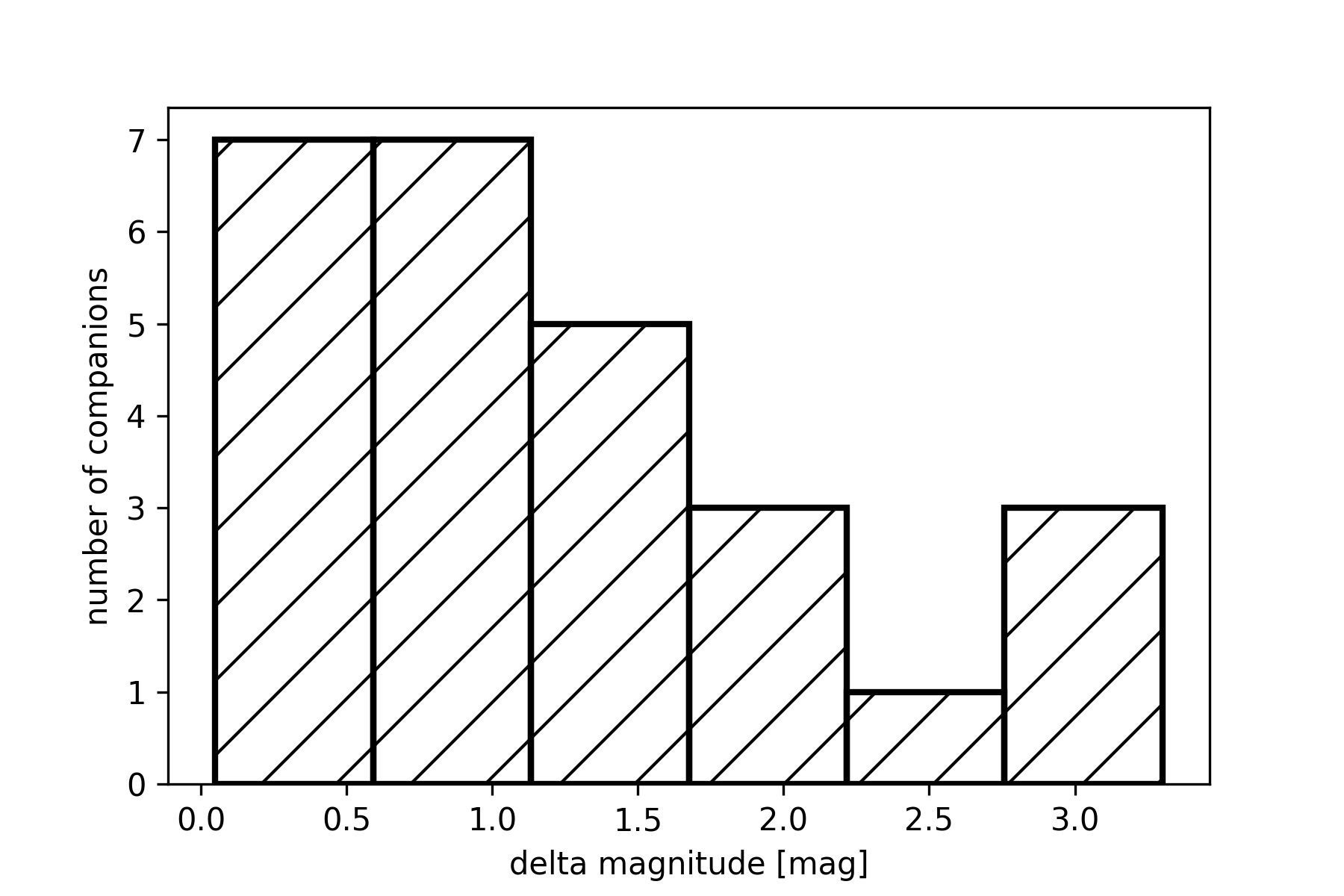}
    \caption{Angular separation (left) and I-band or ``$>700$ nm'' delta magnitude (right) distributions for the \newcompanions{} new discoveries revealed by the POKEMON survey. These values were obtained from our DSSI and NESSI speckle observations, and have not been corrected for anisoplanatism.}
    \label{fig:observed_properties_distributions}
\end{figure*}

\begin{figure*}
    \centering
    \includegraphics[width=\textwidth]{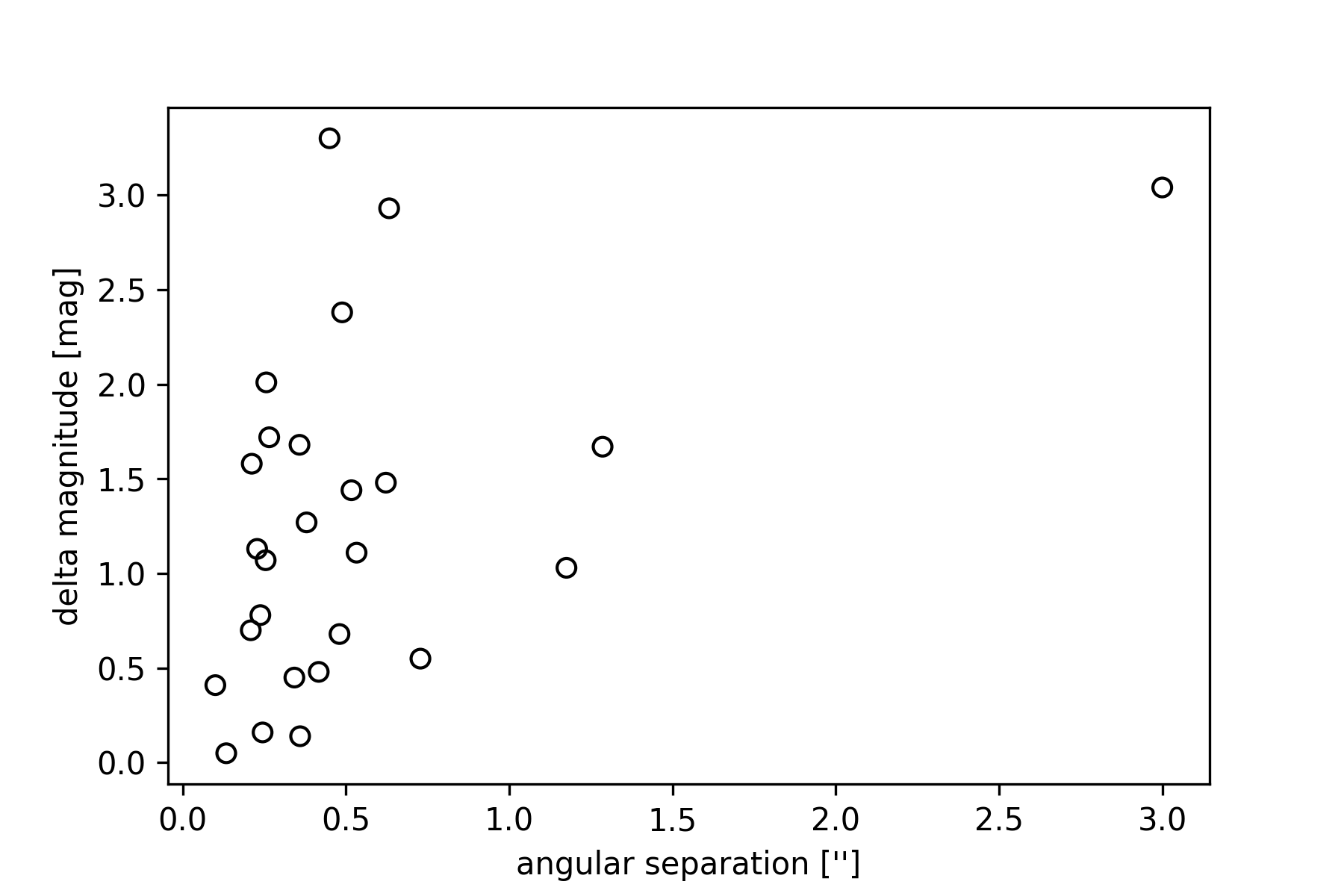}
    \caption{Scatter plot of delta magnitude versus angular separation for the \newcompanions{} new discoveries. Most new discoveries are measured to be within $\sim0.7\arcsec$ of their primaries, though they span a range of delta magnitudes.}
    \label{fig:delta_mag_vs_angular_sep}
\end{figure*}

The observed properties for the new companions are reported in Table \ref{table:observed_properties}, where we have included the 2MASS ID of the primary, date of observation (measured in Besselian years), seeing, position angle ($\theta$), angular separation ($\rho$), delta magnitude ($\Delta$m), and bandpass ($\lambda$). Again, for filterless observations, instead of listing the bandpass of the observation, we instead indicate whether the companion was detected in the ``blue'' ($\lambda\lessapprox700$ nm) image or the ``red'' ($\lambda\gtrapprox700$ nm) image. We also include entries for the known companion to 2MASS J12190600+3150433, since it was detected in our observations as well. The angular separation and delta magnitude distributions for the new discoveries are shown in Figure \ref{fig:observed_properties_distributions}. A scatter plot of delta magnitude versus angular separation is shown in Figure \ref{fig:delta_mag_vs_angular_sep}.

\begin{deluxetable*}{cccccrr}
\tablecaption{Observed properties for systems with a new companion
\label{table:observed_properties}}
\tablehead{\colhead{2MASS ID} & \colhead{Date} & \colhead{Seeing} & \colhead{$\theta$} & \colhead{$\rho$} & \colhead{$\Delta$m} & \colhead{$\lambda$} \\ 
\colhead{} & \colhead{(2000+)} & \colhead{($\arcsec$)} & \colhead{($^{\circ}$)} & \colhead{($\arcsec$)} & \colhead{} & \colhead{(nm)}}
\startdata
 $03104962-1549408$ & 17.8028 & 0.86 & 281.3 & 0.3577 & 1.68 & 880 \\ 
 $03542561-0909316$ & 17.8055 & 1.23 & 39.5 & 2.9997 & 3.04 & 880 \\
 $07011725+1348085$ & 17.2874 & 1.1 & 93.7 & 1.2858 & 1.67 & red \\
 $07411976+6718444$ & 17.2929 & 0.86 & 200.7 & 0.2085 & 0.70 & red \\
 $09510964-1219478$ & 17.3477 & 1.73 & 38.5 & 0.4497 & 3.30 & 880 \\
 $11030845+1517518$ & 17.2931 & 0.72 & 48.8 & 0.3912 & 1.42 & blue \\
 & 17.2931 & 0.81 & 48.1 & 0.3908 & 1.54 & red \\
 & 18.0851 & 0.70 & 44.4 & 0.3694 & 1.53 & 692 \\
 & 18.0851 & 0.68 & 43.9 & 0.3680 & 1.00 & 880 \\
 $12190600+3150433$ & 17.2931 & 0.66 & 331.8 & 0.1458 & 0.18 & blue \\
 & 17.2931 & 0.72 & 331.3 & 0.1415 & 0.09 & red \\
 & 18.0851 & 0.61 & 297.3 & 0.1287 & 0.01 & 692 \\
 & 18.0851 & 0.57 & 298.5 & 0.1253 & 0.01 & 880 \\
 & 17.2931 & 0.66 & 218.8 & 1.7697 & 5.54 & blue \\
 & 17.2931 & 0.72 & 222.9 & 1.6789 & 2.3 & red \\
 & 18.0851 & 0.61 & 216.5 & 1.6852 & 2.64 & 692 \\
 & 18.0851 & 0.57 & 218.7 & 1.6814 & 3.72 & 880 \\
 $12435889-1614351$ & 17.2876 & 0.91 & 284.7 & 0.3420 & 0.45 & red \\
 $13092185-2330350$ & 17.2906 & 0.63 & 105.1 & 1.1947 & 0.99 & blue \\
 & 17.2906 & 1.03 & 102.5 & 1.1752 & 1.03 & red \\
 $14235017-1646116$ & 17.2935 & 0.78 & 18.7 & 0.6604 & 3.29 & red \\
 & 18.0852 & 0.63 & 289.2 & 0.6039 & 2.57 & 880 \\
 $15020759+7527526$ & 17.2825 & 0.56 & 130.5 & 0.6188 & 2.39 & red \\
 & 19.7023 & 0.73 & 135.7 & 0.6256 & 0.56 & 880 \\
 $15085332+4934062$ & 17.2826 & 0.56 & 310.2 & 0.1574 & 1.50 & red \\
 & 20.1085 & 1.34 & 336.5 & 0.2655 & 1.66 & 880 \\
 $15211607+3945164$ & 17.2662 & 0.86 & 90.6 & 0.2378 & 0.78 & red \\
 $15263317+5522206$ & 17.2664 & 0.73 & 218.3 & 0.2280 & 1.13 & red \\
 $15434848+2552376$ & 17.2799 & 0.71 & 262.4 & 0.2448 & 0.31 & blue \\
 & 17.2799 & 0.75 & 262.0 & 0.2447 & 0.16 & red \\
 $15471513+0149218$ & 17.2881 & 0.73 & 287.0 & 0.2650 & 1.72 & red \\
 $16041322+2331386$ & 17.2800 & 0.63 & 109.0 & 0.1065 & 0.85 & blue \\
 & 17.2800 & 0.61 & 107.7 & 0.0982 & 0.37 & red \\
 & 17.3401 & 0.75 & 112.5 & 0.1018 & 0.45 & red \\
 $17183572+0156433$ & 17.2719 & 1.32 & 55.6 & 0.5307 & 1.05 & blue \\
 & 17.2719 & 1.28 & 55.0 & 0.5325 & 1.11 & red \\
 $17335314+1655129$ & 17.2938 & 0.69 & 101.0 & 0.1443 & 1.06 & red \\
 & 19.6995 & 1.05 & 62.6 & 0.3639 & 1.08 & 880 \\
 $18191622-0734518$ & 17.2775 & 0.74 & 132.9 & 0.5168 & 1.44 & red \\
 $18523373+4538317$ & 17.2802 & 0.81 & 37.7 & 0.4920 & 0.54 & red \\
 & 19.6995 & 0.99 & 28.7 & 0.4684 & 0.81 & 880 \\
 $20081786+3318122$ & 19.7026 & 0.69 & 237.6 & 0.4166 & 0.48 & 880 \\
 $21011610+3314328$ & 17.8048 & 1.39 & 348.7 & 0.2649 & 2.07 & 692 \\
 & 17.8048 & 1.19 & 347.8 & 0.2561 & 2.01 & 880 \\
 & 18.5853 & 1.15 & 355.2 & 0.2869 & 1.65 & 832 \\
 $21134479+3634517$ & 18.6600 & 0.65 & 10.6 & 0.3868 & 0.31 & 832 \\
 & 19.6995 & 1.15 & 5.5 & 0.3596 & 0.14 & 880 \\
 $22520522-1532511$ & 19.7027 & 0.60 & 252.4 & 0.7280 & 0.55 & 880 \\
 $23024353+7505591$ & 19.0532 & 0.81 & 3.0 & 0.4862 & 3.30 & 562 \\
 & 19.0532 & 0.63 & 3.7 & 0.4883 & 2.38 & 832 \\
 \enddata
\end{deluxetable*}

We calculate residuals by computing the difference between the observed properties in each channel. The angular separation residuals have an average value of 3.4 mas, with a standard deviation of 6.5 mas. The position angle residuals have an average value of 0.6 degrees, with a standard deviation of 1.0 degrees. The delta magnitude residuals have an average value of 0.06, with a standard deviation of 0.23. The residuals result from the subtraction of two independent measurements with presumably the same uncertainty, so the subtraction has an uncertainty that is $\sqrt 2$ larger than the uncertainty of either individual measure. This means that the uncertainty in a single angular separation measure is given by 6.5 mas divided by $\sqrt 2$, or 4.6 mas. The average uncertainty in position angle is then 0.7 degrees, and the average uncertainty in delta magnitude is 0.16. When we take these values and add them in quadrature to the calibration uncertainties from \citet{Horch2021AJ....161..295H}, we obtain values of 5.1 mas for the average angular separation uncertainty, 2 degrees for the average position angle uncertainty, and 0.21 for the average delta magnitude uncertainty. These uncertainties were propagated throughout our analysis described in Section \ref{subsec:estimated_properties}. These values are slightly larger than those derived in \citet{Horch2017AJ....153..212H}, \citet{Colton2021AJ....161...21C}, or \citet{Horch2021AJ....161..295H}; this is likely due to the faintless of our targets, and the filterless observations that took place during the April 2017 observing run. In general, filterless observations reduce the precision of our astrometry, but allow us to observe fainter companions.

\subsection{Likelihood That the New Discoveries Are Bound} \label{subsec:likelihood}

We note that it is highly unlikely these new discoveries are background stars due to the small fields-of-view of the instruments. We find that the number of stars brighter than $I=15.5$ within a $2\arcsec$ area on the sky is $\sim0.0008$. 
This means that for our \newcompanions{} detections, there is a $\sim2.1\%$ chance that even a single detection is a random background star. Nonetheless, we investigate whether the new discoveries are bound in various ways.

First, \citet{Winters2019AJ....157..216W} list two of our targets found to host a new companion as suspected multiples in their work. In the case of the first suspected multiple, 2MASS J09510964-1219478, no astrometric parameters could be determined in the Double and Multiple Systems Annex \citep{Lindegren1997A&A...323L..53L}, which could indicate that the star is actually a short-period astrometric binary. In the case of the second suspected multiple, \citet{Winters2019AJ....157..216W} used photometry from October 2013 to determine that 2MASS J20081786+3318122 is overluminous. We detected a companion to 2MASS J20081786+3318122 in 2019; there are therefore six years between the epochs of observation, which indicates that the object inducing this overluminosity is a common proper motion companion rather than a background contaminant.

We also assessed the boundedness of the new discoveries using second epoch observations. Because the targets in the POKEMON sample are nearby, they have high proper motions. Therefore, a second epoch observation can be taken to test the new discoveries for common proper motion with the target star. Ten of the \newcompanions{} new discoveries have second epoch observations.

Additionally, we used the 2MASS identifiers of the targets found to have a new companion to find their Early Data Release 3 \citep[EDR3;][]{Gaia2020yCat.1350....0G} identifiers, or in one case their Data Release 2 \citep[DR2; ][]{Gaia2018yCat.1345....0G} identifier, on the Gaia archive\footnote{\url{https://gea.esac.esa.int/archive}}. If a secondary object was listed within $5\arcsec$ of the target, we calculated its angular separation from the target using its coordinates in Gaia. If the angular separation agreed with the separation we measured from our speckle observations, then we examined whether the parallaxes and proper motions of the secondary agreed with those of the primary to within $3\sigma$. We determined that three of our new discoveries are bound using this methodology.

Moreover, \otherparallax{} of our \newcompanions{} systems do not have parallaxes or proper motions listed in Gaia. This often occurs for close-in binary systems, as Gaia is unable to resolve the system, and the relative motion of the pair disrupts the parallax measurements. The absence of Gaia parallaxes or proper motions for these targets therefore suggests that the new discoveries we detected in these nine systems are indeed real.

Furthermore, \citet{Vrijmoet2020AJ....160..215V} notes that three out of four unresolved multi-star red dwarf systems within 25 pc in Gaia DR2 have a parallax error larger than or equal to 0.32 mas, and parallaxes more than $\sim10\%$ different than the long-term RECONS results. Thus, the large Gaia parallax errors for five of the targets found to have a new companion could be caused by the previously unresolved companions that we detected.

Finally, we examined the Gaia EDR3 re-normalized unit weight error (RUWE) values associated with the targets found to have a new companion. The Gaia RUWE metric acts like a reduced chi-squared, where large values can indicate a poor model fit to the astrometry, assuming that the star is single. Single sources typically have RUWE values of $\sim1$, while sources with RUWE values $>1.4$ are likely non-single or otherwise extended \citep{Ziegler2020AJ....159...19Z, Gaia2021A&A...649A...1G}. Following \citet{Vrijmoet2020AJ....160..215V}, which surveyed M dwarfs specifically, we use RUWE $>2$ to distinguish single and (potentially) non-single sources, and find eight targets with elevated RUWE values.

In Table \ref{table:parallax_proper_motion}, we list the parallaxes and proper motions, as well as their errors and sources, for all targets found to have a new companion. We also include entries for the three new discoveries that appear in Gaia. \gaiaparallax{} of the targets found to have a new companion have Gaia parallax estimates either from DR2 or EDR3. Parallaxes for the other \otherparallax{} primaries were obtained using the Fourth Edition of the General Catalogue of Trigonometric Stellar Parallaxes \citep[YPC;][]{vanAltena1995gcts.book.....V}, the MEarth survey \citep{Dittmann2014ApJ...784..156D}, and the URAT Parallax Catalog \citep{Finch2018AJ....155..176F}. Proper motions were obtained either from Gaia or the fourth US Naval Observatory CCD Astrograph Catalog \citep[UCAC4;][]{Zacharias2012yCat.1322....0Z}. We use $*$ to note suspected multiples from \citet{Winters2019AJ....157..216W}. We use \textdagger{} to note targets that have a second epoch observation. We use \textdaggerdbl{} to note the new discoveries that appear in Gaia. We use $\S$ to note targets that do not have parallaxes or proper motions in Gaia EDR3. We use $\|$ to note targets with a Gaia parallax error $>0.32$ mas. Finally, we use \# to note targets with RUWE $>2$.

\begin{longrotatetable}
\begin{deluxetable}{c c c c c c c c}
\tablecaption{Parallaxes and proper motions
\label{table:parallax_proper_motion}}
\tablehead{\colhead{2MASS ID} & \colhead{Parallax} & \colhead{Parallax Error} & \colhead{Parallax Source} & \colhead{Proper Motion} & \colhead{Proper Motion Error} & \colhead{Proper Motion Source} & \colhead{RUWE} \\ 
\colhead{} & \colhead{(mas)} & \colhead{(mas)} & \colhead{} & \colhead{(mas/yr)} & \colhead{(mas/yr)} & \colhead{} & \colhead{}}
\startdata
 $03104962-1549408$ \tablenotemark{$\|$, $\!$\#} & 23.2 & 0.7 & Gaia EDR3 & 325.0, 0.9 & 0.6, 0.7 & Gaia EDR3 & 12.8 \\ 
 $03542561-0909316$ \tablenotemark{\textdaggerdbl} & 47.41 & 0.02 & Gaia EDR3 & -95.41, 111.01 & 0.02, 0.02 & Gaia EDR3 & 1.3 \\
 & 47.54 & 0.03 & Gaia EDR3 & -96.52, 99.07 & 0.03, 0.03 & Gaia EDR3 & 1.9 \\
 $07011725+1348085$ \tablenotemark{\textdaggerdbl} & 24.52 & 0.04 & Gaia EDR3 & 415.50, -97.40 & 0.05, 0.04 & Gaia EDR3 & 1.1 \\
 & 24.1 & 0.1 & Gaia EDR3 & 407.4, -96.4 & 0.2, 0.1 & Gaia EDR3 & 2.8 \\
 $07411976+6718444$ \tablenotemark{$\!$\#} & 41.5 & 0.3 & Gaia EDR3 & -110.3, -273.4 & 0.1, 0.2 & Gaia EDR3 & 9.1 \\
 $09510964-1219478$ \tablenotemark{*} & 76.16 & 0.02 & Gaia EDR3 & 1137.72, -1455.38 & 0.02, 0.01 & Gaia EDR3 & 1.1 \\
 $11030845+1517518$ \tablenotemark{\textdagger, \S} & 55 & 8 & URAT & -419, -84 & 8, 8 & UCAC4 & \\
 $12190600+3150433$ \tablenotemark{\textdagger} & 35.20 & 0.08 & Gaia EDR3 & -295.61, 5.13 & 0.07, 0.08 & Gaia EDR3 & 1.2 \\
 $12435889-1614351$ \tablenotemark{\S} & 51 & 3 & URAT & -422, 74 & 8, 8 & UCAC4 & \\
 $13092185-2330350$ & 66.60 & 0.1 & Gaia EDR3 & 15.7, -383.77 & 0.1, 0.07 & Gaia EDR3 & 1.0 \\
 $14235017-1646116$ \tablenotemark{\textdagger} & 16.93 & 0.03 & Gaia EDR3 & -108.81, -85.83 & 0.03, 0.02 & Gaia EDR3 & 1.5 \\
 $15020759+7527526$ \tablenotemark{\textdagger, \S} & 58 & 3 & MEarth & -132, 160 & 8, 8 & UCAC4 & \\
 $15085332+4934062$ \tablenotemark{\textdagger, $\!$\#} & 25.2 & 0.3 & Gaia EDR3 & -104.9, -13.3 & 0.3, 0.4 & Gaia EDR3 & 27 \\
 $15211607+3945164$ \tablenotemark{\S} & 45 & 1 & MEarth & -436, 178 & 8, 8 & UCAC4 & \\
 $15263317+5522206$ \tablenotemark{\S} & 42 & 4 & MEarth & -111, 235 & 8, 8 & UCAC4 & \\
 $15434848+2552376$ \tablenotemark{\S} & 45 & 2 & MEarth & -171, 317 & 8, 8 & UCAC4 & \\
 $15471513+0149218$ \tablenotemark{$\|$, $\!$\#} & 56.4 & 0.4 & Gaia EDR3 & -213.4, -65.4 & 0.4, 0.3 & Gaia EDR3 & 10 \\
 $16041322+2331386$ \tablenotemark{\textdagger, $\!$\#} & 47.7 & 0.2 & Gaia EDR3 & -162.3, 16.9 & 0.1, 0.2 & Gaia EDR3 & 9.8 \\
 $17183572+0156433$ \tablenotemark{$\|$, $\!$\#} & 27.8 & 0.4 & Gaia EDR3 & -447.5, -283.4 & 0.4, 0.3 & Gaia EDR3 & 17 \\
 $17335314+1655129$ \tablenotemark{\textdagger, $\|$, $\!$\#} & 60.9 & 0.5 & Gaia EDR3 & -135.0, -130.5 & 0.5, 0.4 & Gaia EDR3 & 16 \\
 $18191622-0734518$ \tablenotemark{\S} & 10.4 & 0.3 & Gaia DR2 & -175.1, -213.5 & 0.6, 0.5 & Gaia DR2 & \\
 $18523373+4538317$ \tablenotemark{\textdagger, \S} & 46 & 2 & MEarth & 206, 465 & 8, 8 & UCAC4 & \\
 $20081786+3318122$ \tablenotemark{*, \S} & 46 & 5 & YPC & 340, 375 & 8, 8 & UCAC4 & \\
 $21011610+3314328$ \tablenotemark{\textdagger, $\|$, $\!$\#} & 49.8 & 0.7 & Gaia EDR3 & 325.6, -164.2 & 0.6, 0.7 & Gaia EDR3 & 37 \\
 $21134479+3634517$ \tablenotemark{\textdagger, \S} & 51 & 2 & URAT & -15.7, -92.3 & 6.5, 6.9 & UCAC4 & \\
 $22520522-1532511$ & 37.64 & 0.07 & Gaia EDR3 & 331.41, 14.47 & 0.08, 0.07 & Gaia EDR3 & 1.1 \\
 $23024353+7505591$ \tablenotemark{\textdaggerdbl} & 18.88 & 0.01 & Gaia EDR3 & 285.23, 22.88 & 0.02, 0.02 & Gaia EDR3 & 1.1 \\
 & 18.91 & 0.01 & Gaia EDR3 & 286.33, 17.36 & 0.01, 0.02 & Gaia EDR3 & 1.2 \\
 \enddata
 \tablenotetext{*}{ \citet{Winters2019AJ....157..216W} suspected multiple}
\tablenotetext{$\textdagger$}{ Second epoch observation}
\tablenotetext{$\textdaggerdbl$}{ New discovery in Gaia}
\tablenotetext{\S}{ No Gaia EDR3 parallax or proper motions}
\tablenotetext{\|}{ Gaia parallax error $>0.32$ mas}
\tablenotetext{\#}{ RUWE $>2$}
\end{deluxetable}
\end{longrotatetable}

The only stars that do not have an indicator that they are non-single are 2MASS J13092185-2330350 and 2MASS J22520522-1532511. 2MASS J13092185-2330350 is a unique case where we obtained 46 data cubes on the same target because of its faintness, which could explain why its small and faint companion did not induce one of these indicators; this system is discussed further in Section \ref{subsec:brown_dwarf}. In the case of 2MASS J22520522-1532511, a secondary object does appear in Gaia EDR3, but it does not have a parallax or proper motions listed; this secondary object could potentially be the new discovery we detected.

\subsection{Estimated Astrophysical Properties} \label{subsec:estimated_properties}

Here we estimate projected separations, component masses and system mass ratios, and component spectral types for the \newcompanions{} systems with new discoveries, assuming that there are no additional unknown components in each system (e.g., spectroscopic components).

We calculated projected separations for these \newcompanions{} systems using our measured angular separations and the parallax estimates given in Table \ref{table:parallax_proper_motion}.

We then estimated mass ratios for these \newcompanions{} systems using the empirical relationship between mass and luminosity \citep{Delfosse2000A&A...364..217D, Benedict2016AJ....152..141B}. This relationship was recently calibrated for late-type stars by \citet{Mann2019ApJ...871...63M}. Their code, which is publicly available on $\tt github$\footnote{\url{https://github.com/awmann/M_-M_K-}}, calculates a mass and its uncertainty given a user-provided distance and apparent $K$ magnitude (and uncertainties). We calculated our distances using the parallaxes from Table \ref{table:parallax_proper_motion}, and obtained our $K$ photometry from 2MASS \citep{Skrutskie2006AJ....131.1163S}. However, the resolution for 2MASS is estimated to be 5'', and all of the new companions detected throughout this survey are within 5'' of their primary stars. This means that the 2MASS $K$ magnitudes for these systems convolve the flux from both the primary and secondary stars. In order to separate the 2MASS $K$ magnitudes into their component parts, we first estimated a $K$-band delta magnitude using the method outlined by \citet{Lamman2020AJ....159..139L}. For this we used our measured speckle $I$-band delta magnitude, or the ``$>700$ nm'' delta magnitude in the case of the filterless observations, and the PARSEC theoretical stellar evolution models \citep{Marigo2017ApJ...835...77M}.

We then derived two equations to solve for the $K$ magnitude of the primary and secondary stars. For this we used the $K$-band delta magnitude and the 2MASS $K$ magnitude of the primary star.

The first equation we derived is simply the difference between the secondary $K$ magnitude and the primary $K$ magnitude:

\begin{equation}
\label{eq:delta_K}
    \Delta K = K_2 - K_1
\end{equation}

The second equation originated from the equation for apparent magnitude, which is defined as a logarithmic luminosity ratio of a body to some standard.

\begin{equation}
    m = -2.5 \log_{10}(\frac{L}{L_0})
\end{equation}

The combined apparent magnitude of the binary system can then be written as

\begin{equation}
    m_{binary} = -2.5 \log_{10}(\frac{L_{binary}}{L_0})
\end{equation}

where

\begin{equation}
    \frac{L_{binary}}{L_0} = \frac{L_1}{L_0} + \frac{L_2}{L_0} = 10^{-m_1/2.5} + 10^{-m_2/2.5}
\end{equation}

Our second equation is therefore

\begin{equation}
\label{eq:K_system}
    K_{binary} = -2.5 \log_{10}(10^{-K_1/2.5} + 10^{-K_2/2.5})
\end{equation}

We then used the $\tt scipy.optimize$ subpackage \citep{SciPy2020} to solve Equations \ref{eq:delta_K} and \ref{eq:K_system} for our two unknowns: the $K$ magnitude of the primary and the $K$ magnitude of the secondary.

When observing 2MASS J12190600+3150433, we detected the companion known to the literature in addition to the new discovery. Therefore, to calculate the component $K$ magnitudes for this system, we used an additional equation to calculate the $K$-band delta magnitude for the tertiary in the system, and added a $10^{-K_3/2.5}$ term to Equation \ref{eq:K_system}.

Once we obtained a $K$ magnitude for each component in a given system, we were able to estimate the component masses and calculate a mass ratio for the system. We note that the relationship between mass and luminosity only behaves for Main Sequence stars; in the case of pre-Main Sequence stars, their ages must be known and accounted for as well. The M dwarfs, and in particular the later spectral sub-types, have extensive pre-Main Sequence phases. Therefore, the masses for young members of our sample could be estimated incorrectly if their ages are unknown. \citet{Winters2019AJ....157..216W} finds 17 known young members in their sample of 1120 nearby M dwarfs. Six of these are in our sample as well; however, none of these young stars host a new discovery.

We also determined component spectral types using the component masses we obtained, as well as reference stellar masses and spectral types\footnote{\raggedright``A Modern Mean Dwarf Stellar Color and Effective Temperature Sequence'', \url{http://www.pas.rochester.edu/~emamajek/EEM_dwarf_UBVIJHK_colors_Teff.txt}} \citep{Pecaut2012ApJ...746..154P, Pecaut2013ApJS..208....9P}. We note that this is a coarse method for estimating spectral types; we are therefore currently using the Titan Monitor (TiMo) facility at Lowell Observatory to obtain homogeneous spectral types for the entire POKEMON sample.

The projected separations, mass ratios, and component spectral types for these \newcompanions{} systems are reported in Table \ref{table:estimated_properties}. We also include an entry for the known companion to 2MASS J12190600+3150433, since it was detected in our observations as well.

We note that for some of the \newcompanions{} stars with new companions, multiplying their seeing value by their separation value from Table \ref{table:observed_properties} produces a value larger than 0.6 arcseconds squared. As shown in previous LDT speckle papers \citep{Horch2015AJ....150..151H, Horch2020AJ....159..233H}, a value larger than 0.6 arcseconds squared could indicate that the photometry has a systematic error, which could make the values listed in Table \ref{table:estimated_properties} unreliable. We note these systems with an asterisk.

\begin{longrotatetable}
\begin{deluxetable}{c c c c c c c}
\tablecaption{Estimated astrophysical properties for systems with a new companion
\label{table:estimated_properties}}
\tablehead{\colhead{2MASS ID} & \colhead{Projected Separation} & \colhead{Primary Mass} & \colhead{Companion Mass} & \colhead{Mass Ratio} & \colhead{Primary Spectral Type} & \colhead{Companion Spectral Type} \\ 
\colhead{} & \colhead{(au)} & \colhead{($M_{\odot}$)} & \colhead{($M_{\odot}$)} & \colhead{} & \colhead{} & \colhead{}}
\startdata
 $03104962-1549408$ & $15.4\pm0.6$ & $0.29\pm0.03$ & $0.15\pm0.01$ & $0.53\pm0.09$ & M3 & M5 \\ 
 $03542561-0909316$\tablenotemark{*} & $63.28\pm0.04$ & $0.54\pm0.03$ & $0.19\pm0.02$ & $0.35\pm0.05$ & M0.5 & M4 \\
 $07011725+1348085$\tablenotemark{*} & $52.4\pm0.1$ & $0.30\pm0.03$ & $0.16\pm0.01$ & $0.53\pm0.09$ & M3 & M5 \\
 $07411976+6718444$ & $5.0\pm0.2$ & $0.15\pm0.01$ & $0.120\pm0.008$ & $0.8\pm0.1$ & M5 & M5.5 \\
 $09510964-1219478$\tablenotemark{*} & $5.90\pm0.08$ & $0.50\pm0.03$ & $0.16\pm0.01$ & $0.31\pm0.04$ & M0.5 & M5 \\
 $11030845+1517518$ & $7\pm1$ & $0.28\pm0.05$ & $0.17\pm0.03$ & $0.6\pm0.2$ & M3 & M4.5 \\
 $12190600+3150433$ & $3.8\pm0.3$ & $0.46\pm0.03$ & $0.45\pm0.03$ & $1.0\pm0.1$ & M1.5 & M1.5 \\
 \tablenotemark{*} & $47.7\pm0.1$ & $0.46\pm0.03$ & $0.15\pm0.01$ & $0.33\pm0.05$ & M1.5 & M5 \\
 $12435889-1614351$ & $6.7\pm0.5$ & $0.17\pm0.02$ & $0.15\pm0.02$ & $0.86\pm0.2$ & M4.5 & M5 \\
 $13092185-2330350$\tablenotemark{*} & $17.65\pm0.06$ & $0.082\pm0.004$ & $0.076\pm0.003$ & $0.93\pm0.08$ & M8 & L1 \\
 $14235017-1646116$ & $37.3\pm0.1$ & $0.59\pm0.03$ & $0.23\pm0.02$ & $0.39\pm0.05$ & K8 & M3.5 \\
 $15020759+7527526$ & $10.7\pm0.5$ & $0.114\pm0.007$ & $0.086\pm0.004$ & $0.76\pm0.08$ & M5.5 & M7.5 \\
 $15085332+4934062$ & $8.4\pm0.3$ & $0.51\pm0.03$ & $0.31\pm0.03$ & $0.60\pm0.08$ & M0.5 & M3 \\
 $15211607+3945164$ & $5.3\pm0.3$ & $0.14\pm0.01$ & $0.113\pm0.007$ & $0.8\pm0.1$ & M5 & M5.5 \\
 $15263317+5522206$ & $5.5\pm0.7$ & $0.15\pm0.02$ & $0.11\pm0.01$ & $0.7\pm0.2$ & M5 & M5.5 \\
 $15434848+2552376$ & $5.5\pm0.4$ & $0.16\pm0.01$ & $0.16\pm0.01$ & $0.9\pm0.2$ & M4.5 & M5 \\
 $15471513+0149218$ & $4.9\pm0.2$ & $0.130\pm0.008$ & $0.091\pm0.004$ & $0.70\pm0.08$ & M5 & M6.5 \\
 $16041322+2331386$ & $2.1\pm0.4$ & $0.17\pm0.01$ & $0.15\pm0.01$ & $0.9\pm0.1$ & M4.5 & M5 \\
 $17183572+0156433$\tablenotemark{*} & $19.1\pm0.3$ & $0.35\pm0.03$ & $0.22\pm0.02$ & $0.6\pm0.1$ & M3 & M3.5 \\
 $17335314+1655129$ & $4.2\pm0.2$ & $0.25\pm0.02$ & $0.16\pm0.01$ & $0.7\pm0.1$ & M3.5 & M4.5 \\
 $18191622-0734518$ & $50\pm1$ & $0.52\pm0.03$ & $0.33\pm0.03$ & $0.64\pm0.09$ & M0.5 & M3 \\
 $18523373+4538317$ & $10.4\pm0.6$ & $0.14\pm0.01$ & $0.115\pm0.008$ & $0.8\pm0.1$ & M5 & M5.5 \\
 $20081786+3318122$ & $9\pm1$ & $0.18\pm0.03$ & $0.15\pm0.02$ & $0.8\pm0.2$ & M4.5 & M5 \\
 $21011610+3314328$ & $5.1\pm0.2$ & $0.40\pm0.03$ & $0.18\pm0.02$ & $0.46\pm0.08$ & M2.5 & M4.5 \\
 $21134479+3634517$ & $7.1\pm0.3$ & $0.18\pm0.02$ & $0.17\pm0.02$ & $0.9\pm0.2$ & M4.5 & M4.5 \\
 $22520522-1532511$ & $19.34\pm0.09$ & $0.20\pm0.02$ & $0.16\pm0.01$ & $0.8\pm0.1$ & M4 & M5 \\
 \tablebreak
 $23024353+7505591$ & $25.87\pm0.09$ & $0.67\pm0.02$ & $0.41\pm0.02$ & $0.62\pm0.05$ & K5 & M2 \\
 \enddata
 \tablenotetext{*}{Seeing $\times \rho > 0.6$ arcseconds squared}
\end{deluxetable}
\end{longrotatetable}

We note that our spectral type estimates for 2MASS J14235017-1646116 and 2MASS J23024353+7505591 are K8 and K5, respectively. \citet{Kirkpatrick2016ApJS..224...36K}
lists 2MASS J14235017-1646116 as a M0.5 star, but 2MASS J23024353+7505591 does not have a spectral type listed in the literature. In order to confirm our spectral type estimates, one would need to obtain low-resolution spectroscopy of these objects
, which is outside the scope of this work. However, we are in the process of following up POKEMON targets with the TiMo facility at Lowell Observatory.

\section{Discussion} \label{sec:discussion}

Here we discuss a new companion in a system with both a known M dwarf and a known white dwarf, a potential brown dwarf binary, and initial insights from the POKEMON survey.

\subsection{A Triple System with a Known White Dwarf} \label{subsec:white_dwarf}

As noted in Section \ref{subsec:new_discoveries}, four of the \newcompanions{} systems with new discoveries also host companions known to the literature. The companion to 2MASS J17183572+0156433, called Wolf 672 A, is a known white dwarf \citep{Gianninas2011ApJ...743..138G}.

This system was excluded from the final POKEMON sample, since all systems with primaries more massive than M dwarfs have now been removed. We also do not include this system in our calculation of the M-dwarf multiplicity or companion rates. Nonetheless, we note here the discovery of a new M-dwarf companion in this system, which makes the system trinary.

Systems that include both an M dwarf and a white dwarf are critical to our understanding of M dwarfs, as our low-mass neighbors are notoriously difficult to model, whereas the white dwarfs have a well-defined chain of models \citep{Fontaine2001PASP..113..409F} that can be used to estimate the age of their Main Sequence companion \citep{Garces2011A&A...531A...7G, Kiman2021AJ....161..277K}. This estimation of ages using white dwarfs can have uncertainties as low as 10-20\% \citep{Fouesneau2019ApJ...870....9F}. White dwarf/M dwarf systems can also be used to study mass loss \citep{Debes2006ApJ...652..636D}. This system is therefore of particular interest for understanding the formation and evolution of the M-dwarf companions.

\begin{figure*}
    \centering
    \includegraphics[width=\textwidth]{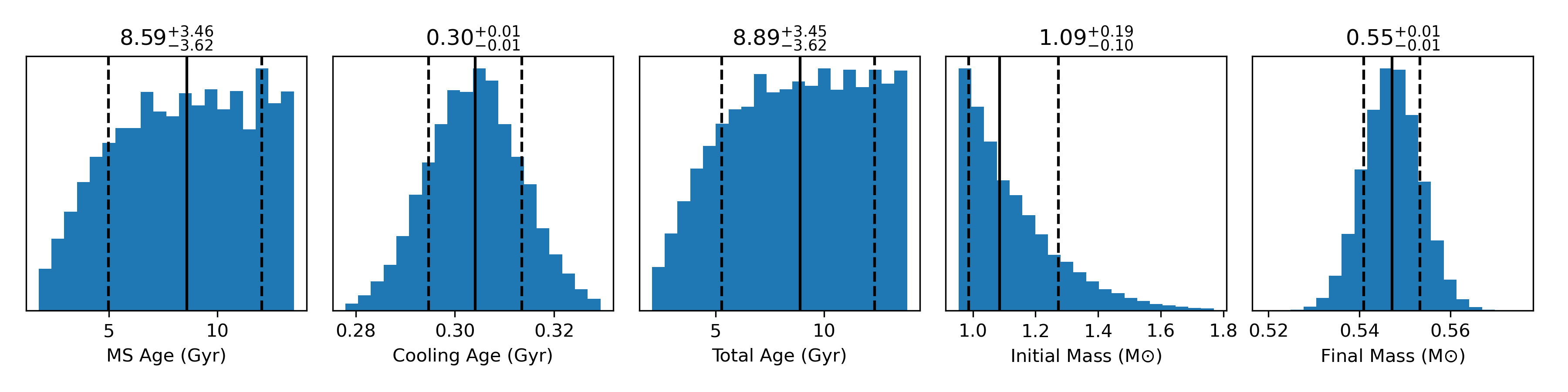}
    \caption{Using the $\tt wdwarfdate$ code, we estimate an age for the white dwarf Wolf 672 A, and thus its companions, the M dwarf Wolf 672 B and the new discovery we detected. The panels (from left to right) show the Main Sequence age (the age of the progenitor), the cooling age, the total age, the initial mass (the mass of the progenitor), and the final mass of the white dwarf. We find a cooling age of $0.30^{+0.01}_{-0.01}$ Gyr, and a final mass of $0.55^{+0.01}_{-0.01} M_{\odot}$, which are in complete agreement with the Montreal White Dwarf Database values for Wolf 672 A of 0.29 Gyr and 0.55 $M_{\odot}$.}
    \label{fig:white_dwarf_age}
\end{figure*}

Wolf 672 A is included in the Montreal White Dwarf Database\footnote{\url{https://www.montrealwhitedwarfdatabase.org}}, where its effective temperature and log $g$ are listed as $12461\pm113$ and $7.89\pm0.01$, respectively. Using these parameters, the fact that the Montreal White Dwarf Database lists Wolf 672 A as a dA white dwarf, and the $\tt wdwarfdate$ code \citep[][Kiman et al. in prep]{Dotter2016ApJS..222....8D, Choi2016ApJ...823..102C, Cummings2018ApJ...866...21C, Bedard2020ApJ...901...93B} that is publicly available on $\tt github$\footnote{\url{https://github.com/rkiman/wdwarfdate}}, we estimate an age for the white dwarf Wolf 672 A, and thus its companions, the M dwarf Wolf 672 B and the new discovery we detected (Figure \ref{fig:white_dwarf_age}). We find an age for this system of $8.89^{+3.45}_{-3.62}$ Gyr, demonstrating the utility of observing white dwarf/M dwarf systems. 

\subsection{A Potential Brown Dwarf Companion} \label{subsec:brown_dwarf}

As noted in Section \ref{subsec:observational_routine}, during our observing run in April 2017, the narrowband speckle filters were not installed within DSSI. Although these filters produce more contrast in the speckles, a large portion of the photons gathered by the telescope remain unused. Filterless observing could thus be beneficial for probing new speckle imaging discovery space, such as in the case of 2MASS J13092185-2330350.

2MASS J13092185-2330350 is listed as an M7 star \citep{Gizis2002ApJ...575..484G}, and has a magnitude of 18.0 in the R band \citep{Reid2007AJ....133.2825R}. On UT April 16 2017, we obtained 46 data cubes for 2MASS J13092185-2330350, rather than the standard nine, using DSSI at the 4.3-meter LDT. These observations were taken at the end of the night, which was an opportunity to test the possibility of observing systems beyond the faint limit of DSSI. As shown in Section \ref{subsec:new_discoveries}, these observations revealed a companion in both channels, at $1.1947\arcsec$ and with a delta magnitude of 0.99 at 692 nm, and at $1.1752\arcsec$ and with a delta magnitude of 1.03 at 880 nm. According to our analysis described in Section \ref{subsec:estimated_properties}, the projected separation for this new companion is 17.64 au. The mass of the primary is 0.082 $M_{\odot}$, and the mass of the secondary is 0.076 $M_{\odot}$, corresponding to a mass ratio of 0.926. We estimate that the spectral type of the primary is M8, and that the spectral type of the secondary is L1, making the secondary a potential brown dwarf. In order to confirm these spectral types, one would need to obtain low-resolution spectroscopy of these objects
; this is outside the scope of this work, though we are currently following up POKEMON targets with the TiMo facility at Lowell Observatory. In any case, this system provides an argument for further investigation into filterless speckle observations, as well as systems beyond the faint limit of the instruments.

\subsection{Initial Insights from the POKEMON Survey}

Though the full analysis of the POKEMON sample has yet to be published, we discuss here initial insights that can be made about the POKEMON sample based on the \newcompanions{} newly discovered companions.

As shown in Figure \ref{fig:observed_properties_distributions}, almost all of the angular separations we measured are within $\sim0.7\arcsec$, and the smallest angular separation was measured to be $0.0982\arcsec$. There are also new companions with delta magnitudes as large as 3.30. We note that these distributions have not been completeness-corrected. Therefore close companions at large delta magnitudes would be preferentially missed, since the sensitivity of the instruments decreases with decreasing separation. Additionally, wide companions would be preferentially missed due to the small field-of-views of the instruments, or preferentially excluded since many wide companions have been detected by previous surveys. Nonetheless, the small angular separations and the large range of delta magnitudes of the new discoveries discussed here explain why speckle imaging was needed to detect these \newcompanions{} new companions.

\begin{figure*}
    \centering
    \includegraphics[width=0.49\textwidth]{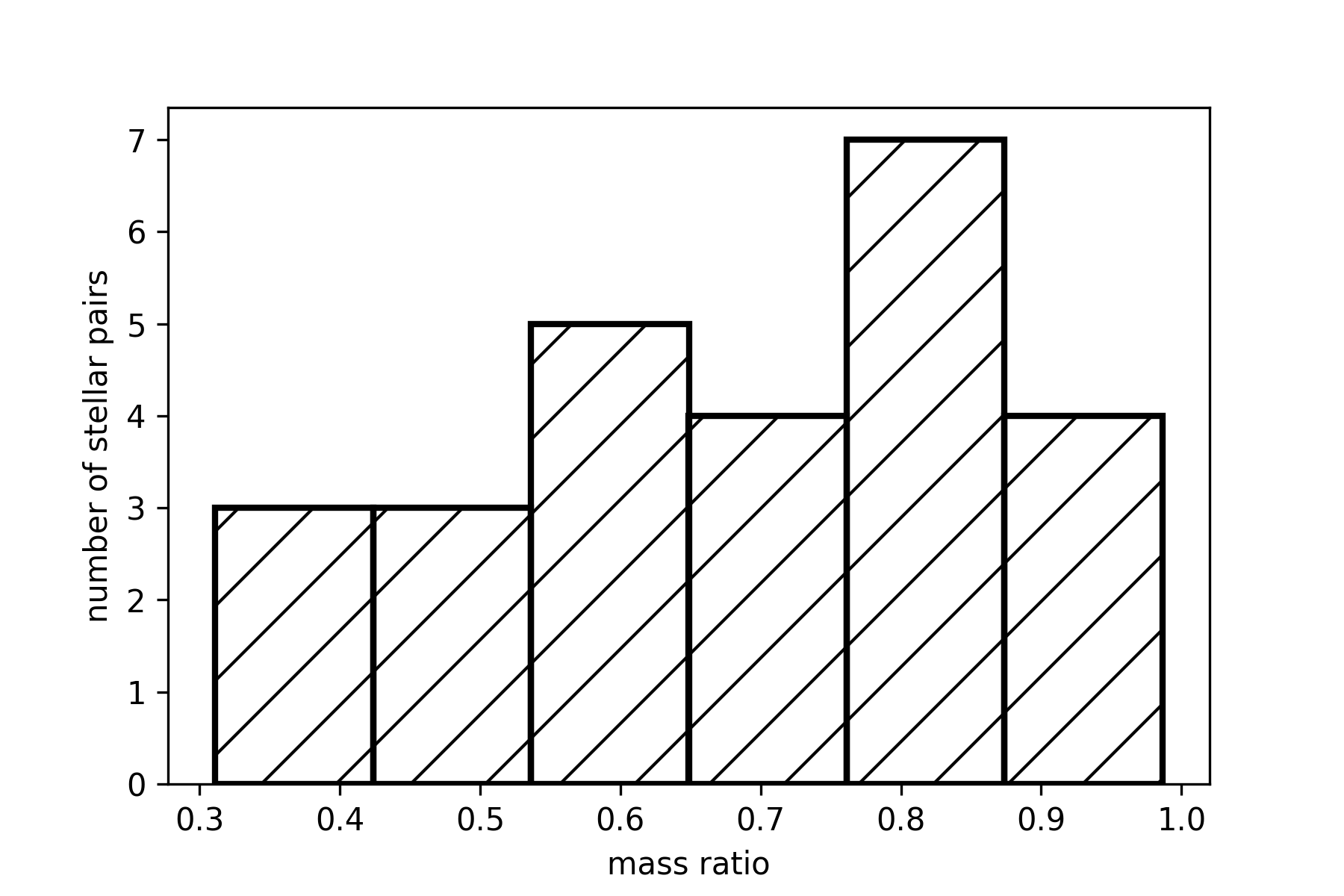}
    \includegraphics[width=0.49\textwidth]{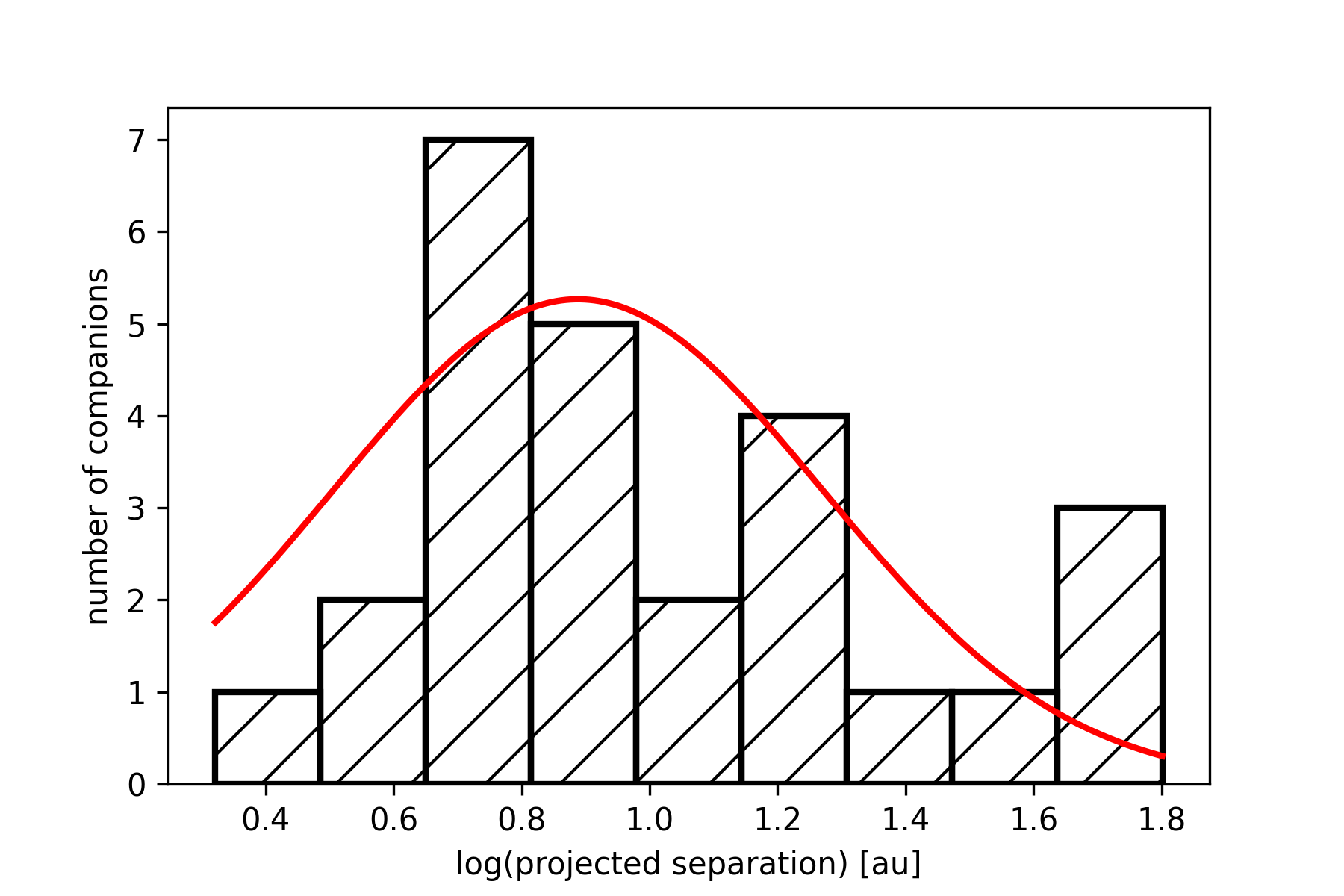}
    \caption{The mass ratio (left) and projected separation (right) distributions for the \newcompanions{} previously undetected companions discovered throughout the POKEMON survey. These values were estimated using our observed properties, as well as the parallaxes and 2MASS $K$ magnitudes for these objects. The mass ratios follow a relatively flat distribution, while the projected separations follow a roughly Gaussian distribution, as expected. The peak of our projected separation distribution is at 7.8 au, with a standard deviation of $\sigma_{\log a} = 0.4$.}
    \label{fig:estimated_properties_distributions}
\end{figure*}

In Figure \ref{fig:estimated_properties_distributions}, we show the mass ratio and projected separation distributions for the \newcompanions{} new companions. The mass ratios follow a relatively flat distribution, while the logarithms of the projected separations follow a roughly Gaussian distribution. The peak of our projected separation distribution is at 7.8 au with a standard deviation of $\sigma_{\log a} = 0.4$. The sample that generated these distributions is of course small; the statistics on these distributions will become more robust once all detections from the full POKEMON survey are included. Nonetheless, our peak is comparable to that of \citet{Winters2019AJ....157..216W}, which found a peak at 20 au for their sample of M dwarfs within 25 pc, and at 4 au for M dwarfs within 10 pc. In contrast, \citet{Raghavan2010ApJS..190....1R} found a peak at 51 au for solar-type stars within 25 pc. It is telling that though we used a different technique than \citet{Winters2019AJ....157..216W} to probe M-dwarf multiplicity, our projected separation distributions are consistent with one another, and are not consistent with the distribution from \citet{Raghavan2010ApJS..190....1R}; these results demonstrate the differences between the stellar companions that M dwarfs host, and those that solar-type stars host.

There are \pokemonnum{} M dwarfs in the POKEMON sample; \knownmultiples{} of these are currently known to host at least one stellar companion. We discovered \newcompanions{} new companions throughout this survey; 22 of these were discovered in systems  that were previously thought to be single. These 22 new discoveries therefore increase the number of M dwarfs in the POKEMON sample with a known companion by \percentincrease{}. Though we have not yet performed sufficient analysis to determine an updated M-dwarf multiplicity rate, these new discoveries increase the companion fraction from 27.1\%, which is in agreement with \citet{Winters2019AJ....157..216W}, to 29.2\%. This increase demonstrates that it is critical to survey our M-dwarf neighbors with high-resolution imaging in order to search for stellar companions within $2\arcsec$, and to fully understand M-dwarf multiplicity.

\section{Conclusions and Future Work} \label{sec:conclusions}

We have carried out the POKEMON speckle survey of nearby M dwarfs, which is volume-limited through M9, out to at least 15 pc, with additional brighter targets at larger distances. The POKEMON survey has resulted in observations of \pokemonnum{} M dwarfs at large-telescope, diffraction-limited resolution. These observations have revealed \newcompanions{} new companions to these objects.

We used our observed properties, parallaxes from the literature, and the targets' 2MASS $K$ magnitudes to estimate astrophysical properties for these objects including projected separation, mass ratio, and component spectral types.

We report the discovery of a new companion in a system with both a known M dwarf and a known white dwarf, and discuss a new companion that is a potential brown dwarf. We also explore initial insights from the POKEMON survey.

In a forthcoming publication, we will present the full POKEMON survey, as well as an updated M-dwarf multiplicity rate, calculated by spectral sub-type through M9 for the first time. We are also currently carrying out follow-up of the targets in the POKEMON sample in various ways. First, we are using the Titan Monitor facility at Lowell Observatory to obtain homogeneous spectral types for the entire POKEMON sample. Additional follow-up observing is being pursued with the Quad-camera Wavefront-sensing Six-channel Speckle Interferometer \citep[QWSSI;][]{Clark2020SPIE11446E..2AC}, which was commissioned at the LDT in July 2020. While DSSI and NESSI image at two visible wavelengths, QWSSI observes at four channels in the optical, as well as two in the near-infrared, and also includes simultaneous wavefront sensing. Because the M dwarfs emit most strongly in the near-infrared, QWSSI will aid us in measuring the multiplicity of the late-type M dwarfs that were unable to be explored by the previous-generation speckle imagers. This new capability will allow us to build on the POKEMON survey and develop a more complete picture of M-dwarf multiplicity for the later sub-types, and potentially even the L- and T-dwarfs.

\acknowledgments

We thank our anonymous reviewer for their thoughtful assessment. We also thank Frederick Hahne, Zachary Hartman, and Joe Llama for their contributions to and feedback on this manuscript. Finally, we thank the army of TOs at the LDT and the WIYN Telescope for all of their help and insight during our 50 nights of observing.

This research was supported by NSF Grant No.~AST-1616084 and JPL RSA No.~1610345.

These results made use of the Lowell Discovery Telescope at Lowell Observatory. Lowell is a private, non-profit institution dedicated to astrophysical research and public appreciation of astronomy and operates the LDT in partnership with Boston University, the University of Maryland, the University of Toledo, Northern Arizona University and Yale University. Lowell Observatory sits at the base of mountains sacred to tribes throughout the region. We honor their past, present, and future generations, who have lived here for millennia and will forever call this place home.

These results are also based on observations from Kitt Peak National Observatory, the NSF's National Optical-Infrared Astronomy Research Laboratory (NOIRLab Prop. ID: 2018B-0126; PI: C. Clark), which is operated by the Association of Universities for Research in Astronomy (AURA) under a cooperative agreement with the National Science Foundation. Data presented herein were obtained at the WIYN Observatory from telescope time allocated to NN-EXPLORE through the scientific partnership of the National Aeronautics and Space Administration, the National Science Foundation, and the NSF's National Optical-Infrared Astronomy Research Laboratory.

This work presents results from the European Space Agency (ESA) space mission Gaia \citep{https://doi.org/10.26131/irsa12}. Gaia data are being processed by the Gaia Data Processing and Analysis Consortium (DPAC). Funding for the DPAC is provided by national institutions, in particular the institutions participating in the Gaia MultiLateral Agreement (MLA). The Gaia mission website is \url{https://www.cosmos.esa.int/gaia}. The Gaia archive website is \url{https://archives.esac.esa.int/gaia}.

This work has used data products from the Two Micron All Sky Survey \citep{https://doi.org/10.26131/irsa2}, which is a joint project of the University of Massachusetts and the Infrared Processing and Analysis Center at the California Institute of Technology, funded by NASA and NSF.

Information was collected from several additional large database efforts: the Simbad database and the VizieR catalogue access tool, operated at CDS, Strasbourg, France; NASA's Astrophysics Data System; the Washington Double Star Catalog maintained at the US Naval Observatory; and the fourth US Naval Observatory CCD Astrograph Catalog \citep{https://doi.org/10.26131/irsa17}.

%

\vspace{5mm}
\facilities{LDT(DSSI), WIYN(NESSI)}


\software{Astropy \citep{Astropy2013}, IPython \citep{IPython2007}, Matplotlib \citep{Matplotlib2007}, NumPy \citep{NumPy2020}, Pandas \citep{Pandas2010}, SciPy \citep{SciPy2020}}

\clearpage




\bibliographystyle{aasjournal}
\bibliography{references}






\end{document}